\documentclass[conference]{IEEEtran}
\IEEEoverridecommandlockouts
\usepackage{cite}
\usepackage{amsmath,amssymb,amsfonts}
\usepackage{subfigure}
\usepackage{parskip}
\usepackage{float}
\usepackage{algorithmic}
\usepackage{graphicx}
\usepackage{textcomp}
\usepackage{xcolor}
\usepackage{mathtools}
\def\BibTeX{{\rm B\kern-.05em{\sc i\kern-.025em b}\kern-.08em
    T\kern-.1667em\lower.7ex\hbox{E}\kern-.125emX}}
\begin{document}

\title{MDistMult: A Multiple Scoring Functions Model for Link Prediction on Antiviral Drugs Knowledge Graph
}

\author{\IEEEauthorblockN{Weichuan Wang}
\IEEEauthorblockA{\textit{School of Computer Science} \\
\textit{Wuhan University}\\
Wuhan, China \\
tandaocmm@whu.edu.cn}
\and
\IEEEauthorblockN{Zhiwen Xie}
\IEEEauthorblockA{\textit{School of Computer Science} \\
\textit{Wuhan University}\\
Wuhan, China \\
xiezhiwen@whu.edu.cn}
\and
\IEEEauthorblockN{Jin Liu\thanks{\IEEEauthorrefmark{1} Jin Liu is corresponding author.}\IEEEauthorrefmark{1}}
\IEEEauthorblockA{\textit{School of Computer Science} \\
\textit{Wuhan University}\\
Wuhan, China \\
jinliu@whu.edu.cn}
\and
\IEEEauthorblockN{YuCong Duan}
\IEEEauthorblockA{\textit{College of Information Science and Technology} \\
\textit{Hainan University}\\
Haikou, China \\
duanyucong@hotmail.com}
\and
\IEEEauthorblockN{Bo Huang}
\IEEEauthorblockA{\textit{School of Electronic and Electrical Engineering} \\
\textit{Shanghai
University of Engineering Science}\\
Shanghai, China \\
huangbosues@sues.edu.cn}
\and
\IEEEauthorblockN{Junsheng Zhang}
\IEEEauthorblockA{\textit{Institute of Scientific and Technical Information of China} \\
Beijing, China \\
zhangjs@istic.ac.cn}
}

\maketitle

\begin{abstract}
Knowledge graphs (KGs) on COVID-19 have been constructed to accelerate the research process of COVID-19. However, KGs are always incomplete, especially the new constructed COVID-19 KGs. Link prediction task aims to predict missing entities for $(e,r,t)$ or $(h,r,e)$, where $h$ and $t$ are certain entities, $e$ is an entity that needs to be predicted and $r$ is a relation. This task also has the potential to solve COVID-19 related KGs' incomplete problem. Although various knowledge graph embedding (KGE) approaches have been proposed to the link prediction task, these existing methods suffer from the limitation of using a single scoring function, which fails to capture rich features of COVID-19 KGs. In this work, we propose the MDistMult model that leverages multiple scoring functions to extract more features from existing triples. We employ experiments on the CCKS2020 COVID-19 Antiviral Drugs Knowledge Graph (CADKG). The experimental results demonstrate that our MDistMult achieves state-of-the-art performance in link prediction task on the CADKG dataset.
\end{abstract}


\section{Introduction}\label{section1}
COVID-19 is an unpredictable disaster for the whole world. Due to this virus, many people died and fell into poverty, while the communication and trade among countries were blocked. COVID-19 also brought a huge amount of challenges for the world, such as how to explore the structure of the virus and how to design new antiviral drugs. Nowadays, researchers focus on these problems by using machine learning and deep learning techniques. In computing and natural language processing (NLP)\cite{belinkov-glass-2019-analysis} areas, researchers proposed unique methods that using the text content from COVID-19 research papers to push on the process of overcoming difficulties made by COVID-19\cite{michel2020covid}. 

Knowledge graph collects great attention from both industry and academia for its strong ability to store data with structured triples, which provide a huge aid for data mining and some inference tasks. Some researchers builds COVID-19 related knowledge graphs\cite{domingo2020covid}\cite{wise2020covid} with the hope that it can help mining information for the whole of humanity to weather the storm. For instance, some open knowledge graph resource: Kaggle\cite{kaggle}, OpenKG\footnote{http://openkg.cn/dataset/covid-19}, collected COVID-19 related information over 400,000 scholarly articles, including over 150,000 papers with full text, SARS-CoV-2, and related coronaviruses. Link prediction is a common NLP task that it needs to predict entities or relations for an incomplete triple ($(h,r,?)$ or $(?,r,t)$ or $(h,?,t)$) with plenty of existing triples in the knowledge graph. In the fight against the COVID-19, link prediction can show its value on predicting COVID-19 related things, such as patients trajectory, antiviral drugs, etc. In this work, we use the CCKS2020 COVID-19 Antiviral Drugs Knowledge Graph Dataset (CADKG)\footnote{https://www.biendata.xyz/competition/ccks\_2020\_7\_3}as COVID-19 dataset to predict COVID-19 related viruses, proteins and drugs.

Knowledge graph embedding (KGE) models as commonsense methods to complete the link prediction task. In our investigation, KGE models have high performance on some open datasets (e.g. FB15K\cite{bordes2013translating}, FB15K-237\cite{2015Observed}). However, low performance in some domain areas is often observed (e.g. TransE achieved the 0.72 MRR on FB15K but has only a 0.14 MRR on CADKG). In the past few decades, many researchers do plenty of great jobs on link prediction task. Translation-based embedding (TransE\cite{bordes2013translating}) method projects entities and relations as low dimension vectors, and TransH el\cite{2014Knowledge}\cite{article} improved the TransE scoring function to get better performance. Nevertheless, these translation-based model can’t get rid of the inherent limitation\cite{wang2018multi}. Other KGE methods try to promote performance on other parts. Some researchers\cite{nayyeri2019toward} paid their attention to changing loss functions to overcome part of the inherent limitation of TransE. Apart from that, RotatE\cite{2019RotatE} and TorusE\cite{ebisu2018toruse} used lots of negative samplings and set big embedding dimension (d=10000) respectively to improve performance. We measure these above KGE methods, but they all have low-performance phenomena on CADKG. In this work, we pay more attention to improving the amount and types of scoring function, which largely determines the final results on link prediction task. In the past, researchers wanted to use or verify a single scoring function to achieve better results, and it made the model easy to train and could easily apply on large scale datasets. By comparing parameters and results of these models, we find the models can reach better performance when they extract more features on datasets like CADKG. Deploying more scoring functions is the direct way to extract more features, which is worth considering carefully.

In this study, following the above ideas, we propose a new multiple scoring functions model, which called MDistMult (multiple DistMult joint model) model. MDistMult has multiple DistMult scoring functions and can overcome the dilemma that single models met. For a single DistMult model, it just has one relation matrix, but the MDistMult model has multiple relation matrices and the amount of matrices is equal to the number of the DistMult models. On the one hand, MDistMult can extract more features between entities in triple $(h,r,t)$ through multiple scoring functions. On the other hand, it can overcome the existed \emph{symmetry problem} in the single DistMult model, which is defined and complained in section \ref{section3}. We compare our proposed model with common KGE methods. The experimental results show that our method achieves state-of-the-art performance. Compared with other models, the dimension change experiment shows that when the embedding dimension increased, our model can reach better performance. This means that our model can extract more features from our unique design. The main contribution of this work can be summarized as follows:
\begin{itemize}
\item We propose a new multiple scoring functions model named MDistMult for link prediction on the antiviral drug knowledge graph. By designing multiple scoring functions, MDistMult not only extracts more features of existing triples but also solve the \emph{symmetry problem} at the same time.
\item Experimental results show that our model achieves state-of-the-art results on CCKS 2020 antiviral drugs knowledge graph (CADKG) dataset in the link prediction task. Further exploration on the effect of dimension change also illustrates that our model can extract more features on triples when the dimension increases.
\end{itemize}

\section{Related Work}\label{section2}
Link prediction aims at predicting the loss entities in a triple $(h,r,?) $ or $(?,r,t)$ where $h$, $t$, $r$ are the head entity, the tail entity and the relation respectively. Currently main KGE methods focus on predicting the missed entities.
The mainstream research KGE methods learn the entities and relations in triples as low-dimension embedding vectors, then use the designed scoring function or model to calculate the confidence or probability of incomplete triples and predict the missed entities.
KGE methods can be summarized as four types, which can be called 'single model' in our work, including Translation Models, Bilinear Methods, Neural Network Methods and Rotation Models. We describe these models details as follows:
\begin{itemize}
    \item \textbf{Translation Models:} Translation models view the triples as a translation mechanism: the head entity translates to the tail entity by the relation. The most classic method is TransE\cite{bordes2013translating},in which the scoring function is simple as $\textbf{h}+\textbf{r}=\textbf{t}$ for every positive triple. While researchers find the methods can't predict the 1-n, n-1 and n-n relation. Then the TransH\cite{2014Knowledge} was proposed, the TransH made the head entity and tail entity project to hyperplane for predicting n-n relation. Furthermore, in order to consider the different attributes that different relations focus on, the TransR\cite{article} put the entities embedding and relations embedding on different spaces.
    \item \textbf{Bilinear Methods:} Bilinear methods calculate the confidence in vector space between entities and relations. Including RESCAL\cite{2011A}, DistMult\cite{2014Embedding}, ComplEx\cite{2016Complex} and etc. The scoring function of RESCAL is $h^tM_rt$, it sets a non-singular matrix for relation, and plenty of matrix operations to make deeper interaction between entities and relations. But it can reach over-fitting easily, meanwhile, the complexity of it is so high that hard to apply on large scale knowledge graphs. To figure out the problem of RESCAL, DistMult relaxes the restrictions for the relation matrix and changes it to a diagonal matrix. The scoring function of DistMult is $h^Tdiag(r)t$. It is clear that the matrix operation becomes easier so DistMult can be applied on large scale KG. However, the property of the diagonal matrix means that DistMult will regard each relation as a symmetric one in KG. ComplEx expands the DistMult method on complex space and can predict symmetric relation and asymmetric relations concurrently. The thought of using complex spaces also provides huge help on later link prediction methods.
   \item \textbf{Neural Network Methods:} Neural network methods utilise various neural network structures to obtain the interactive information between entities and relations in positive triples. ConvE\cite{2017Convolutional} merges the head embedding and the relation embedding to a new 2-dimension tensor, then gets the mutual information from the convolution filter and full-connection layer, finally having matrix calculate with the entity matrix and using \textsl{Softmax} to calculate the confidence of the input triple. CapsE\cite{2018A} exploits capsule neural network to compute the confidence of positive triples. However, neural network models have a remarkable problem: these models lack convinced explanations for their great performance.
   \item \textbf{Rotation Models:}
   Rotation models regard the relation as a rotation process between entities. The most representative model is RotatE\cite{2019RotatE}. RotatE holds the opinion that amount of types of relations existed in KGs, such as symmetry, anti-symmetry, inversion and composition. RotatE proposes to model on the two-dimensional complex plane space to figure out these relations and treat the relationship as a rotation between \emph{head} and \emph{tail}. QuatE\cite{2019Quaternion} further extends rotation to three-dimensional complex planes and exploits quaternions to realize the rotation operation which avoids the deadlock in the Euler angle as well.
\end{itemize}

We notice that Translation Models, Bilinear Methods and Rotation Models simply use single scoring function to train knowledge graph embedding, but these designed functions limit the models' ability to extract more features, which can eventually enhance the performance. Although Neural Network Methods exploit the properties of CNN to mining as more features as possible, they still suffer from the defect of lack of interpretability which other three methods have. To solve these two problems, we design MDistMult which can extract more features with interpretability to solve both problems.

\section{Methodology} \label{section3}
In this section, we firstly define the link prediction task, positive triples and negative triples, then introduce the single DistMult model, finally describes the MDistMult model and how to build it. Meanwhile, we depict the whole scoring function and the designed loss function. We also give the proof that it can overcome the symmetric prediction problem made by the single DistMult model.

\subsection{Task Definition}
The goal of the link prediction task is to find the true entity that the triple as $(h,r,?)$ or $(?,r,t)$ missed. Some mathematical definitions related to the task are given as follows:
\begin{itemize}
    \item For a given knowledge graph named KG, a triple $t = (e_i,r_m,e_j), e_i,e_j\in{E}, r_m \in R,t \in{T}$ in KG, called a positive triple. The $E$ and $R$ are entities' set and relations' set respectively, and the $T$ is a set of positive triples. The set of negative is $t^{'}= (e_q,r,e_m), e_q,e_m\in{E}, r \in R$, and $t^{'}\notin{T}$, the triple $t^{'}$ is defined as a negative triple, and we define the set of negative triples as $T^{'}$. In the end, we can define the goal of link prediction task is to find the true $e_0$, which is unknown in the triple $(e_0,r,e_i), e_i\in{E}$ or the triple $(e_i,r,e_0), e_i\in{E}$. 
\end{itemize}
\subsection{DistMult Model}
\begin{figure*}[t]
  \begin{center}
  \includegraphics[width=0.7\textwidth]{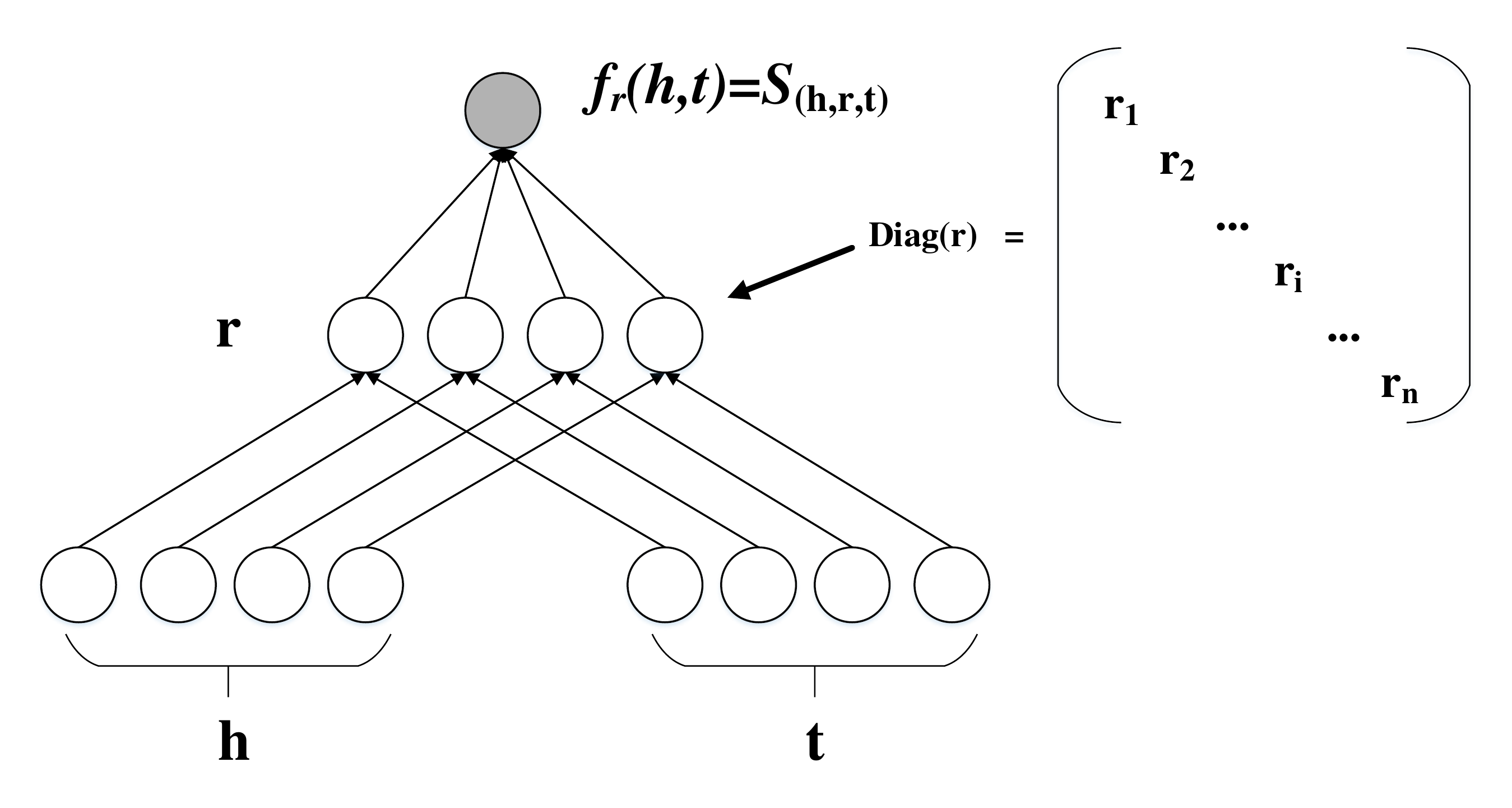}
  \caption{DistMult model structure. The bold $h$,$r$,$t$ are embedding for \emph{head entities}, \emph{relations} and \emph{tail entities} respectively. $r$ is a Diagonal matrix, which is shown in the upper diagram's right part. $f_{r}(h,t)$ is the scoring function of DistMult model, and it can also be written as $S_{(h,r,t)}$. The result of the scoring function means the confidence of the calculated triple. }
  \label{DistMult}
 \end{center}
\end{figure*}
Figure \ref{DistMult} visualizes the DistMult model structure, it consists  of  four  parts:  entity  representations,  relation  representations, scoring function and loss function. We show the  details of these four parts as follows:
\begin{itemize}
    \item \textbf{Entity Representation:} Denote by $x_{e1}$ and $x_{e2}$ the input vectors for entity $e_1$ and $e_2$ respectively. The learned entity representations, $y_{e1}$ and $y_{e2}$ can be written as:
    \begin{equation}
    	\mathbf{y}_{e_1} = f(W\mathbf{x}_{e_1}),\mathbf{y}_{e_2} = f(W\mathbf{x}_{e_2})
    	\label{Disequ1}
\end{equation}
where $f$ can be a linear or non-linear function, and $W$ is a projection parameter matrix, which can be randomly initialized or initialized by using pre-trained matrix.
    \item \textbf{Relation Representation:} The relation representation is a diagonal matrix:
    \begin{equation}
        diag(r_1,r_2,...,r_n)
    \end{equation}
    where $diag()$ means a diagonal matrix, $r_i$ means the parameters of relation representation, which abbreviated as $diag(r)$ and can be randomly initial, the amount of parameters in $diag(r)$ is equal to the length of $y_{e_1}$.
    \item \textbf{Scoring function:}
    The scoring function $f$ of DistMult model is as follow:
    \begin{equation}
        f_{r}(h,t)={S_{(e_1,r,e_2)}}=\mathbf{{y_{e_1}}^T}diag(r)\mathbf{{y_{e_2}}}
    \end{equation}
    where the $\mathbf{y_{e1}}$ and $\mathbf{y_{e2}}$ are head entity and tail entity respectively, which have been mentioned above.
    \item \textbf{Loss Function:} Give a positive triple $t = (e_1,r,e_2),t\in T$, we can replace either the $e_1$ or the $e_2$ to get a negative triple $t^{'} = (e_{1}^{'},r,e_{2}),t^{'}\in T^{'}$ or $t^{'} = (e_{1},r,e_{2}^{'}), t^{'}\in T^{'}$. Denote the scoring function for triple $t = (e_1, r, e_2)$ as $S_{(e_1,r,e_2)}$. The training objective is to minimize the margin-based ranking loss:
    \begin{equation}
    \begin{split}
    L\left ( \Omega \right )=\sum_{(e_{1},r,e_{2})} \sum_{(e_{1}^{'},r,e_{2}^{'})}^{}max \{&S_{(e_{1}^{'},r,e_{2}^{'})}-\\
    &S_{( e_{1},r,e_{2})}+1,0 \}
    \label{Disequ3}
    \end{split}
    \end{equation}
   where 1 means the margin which was set in \cite{2014Embedding} and 0 means dropping out the negative number, and this operation can make the training process shorter.
    
\end{itemize}
\subsection{MDistMult}
\begin{figure*}[t]
  \begin{center}
  \includegraphics[width=0.85\textwidth]{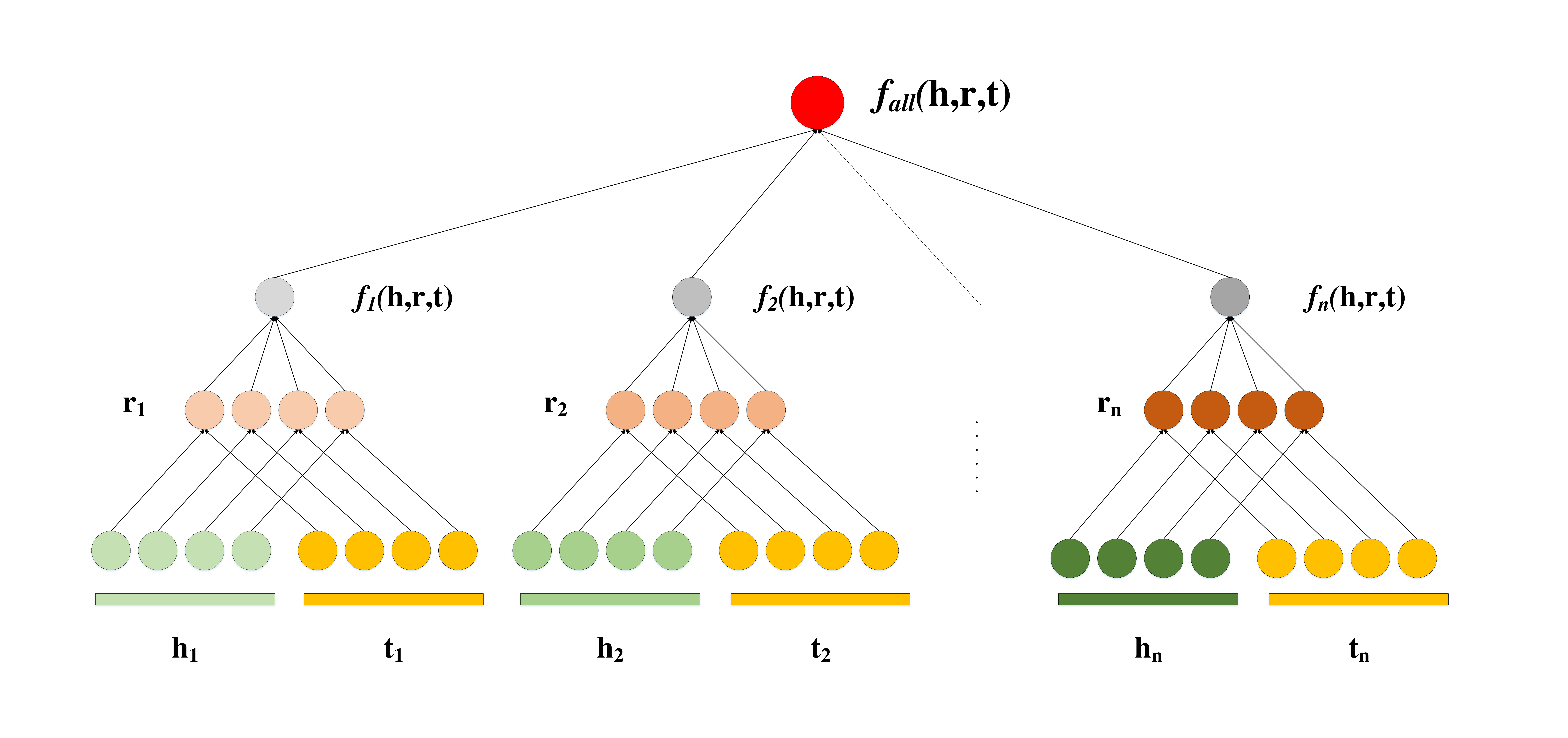}
  \caption{MDistMult model structure. For the $h_{i}$, $t_{i}$, $r_{i}$, circles of different colors represent different embedding. The yellow circles mean the shared tail entities in the whole MDistMult model. The circles with diverse colors as $f_{i}(h,r,t)$ have different parameters in scoring functions. The $f_{all}(h,r,t)$ adds all the scoring functions of single DistMult modules in MDistMult model. All the scoring functions are used in our proposed model, so we call it MDistMult model.}
  \label{MDistMult}
  \end{center}
\end{figure*}
We propose the MDistMult model that improves from the DistMult model as a multiple scoring functions model, which has multiple DistMult's scoring functions. To design the 'Mmodel' which has multiple scoring functions, we need to figure out three important problems: firstly, each 'single model' within the 'Mmodel' must have different parameters, secondly, the ‘Mmodel’ must overcome the disadvantages of original 'single model', thirdly, the joint method is important too, we must design proper loss function to train the 'Mmodel', and the loss functions should be designed for each 'single model' and the whole model. Following the principle of solving these three problems, previous models mainly have two problems:
\begin{itemize}
    \item \textbf{Too Many Parameters:}
    Some models have great performance, but the parameters of them are too many, which make it so hard to expand that be a candidate single model. e.g. Neural network models, RESCAL.
    \item \textbf{Hard to Joint:}
    Translation models have simple scoring functions and are easy to explain, which made them easy to understand. However, when multiple single translation models are bound together. We can ensure that each of its trains is well, which means that it has a local optimum, but it's hard for us to design the whole loss function to ensure it has a global optimum. The rotation models have the same problem. What's more, with the increase of the number of models, the amount of calculation is also increased.
    \end{itemize}
To solve the above two problems, we need to find a simple model as a module of our proposed models, and the module should have two properties:
\begin{itemize}
    \item \textbf{Limited Parameters.} This means the whole designed model's parameters linearly increase when the number of chosen modules rises. In addition, the number of parameters of the module should as small as possible. 
    \item\textbf{Easy to Joint.} The chosen module can provide a shared part and preserve the independent part at the same time. This means each module in designed model can share a part of parameters and have another part of different parameters with other modules.
\end{itemize}

By comparing all the current KGE models, we finally find the DistMult is the most appropriate one. 

The space complexity of DistMult model is $O(N(m+n))$, in which $N$ is the setting dimension, while $m$ and $n$ are the number of entities and relations respectively. Besides, the DistMult model's calculation is simple and easy to combine. Now we just need to solve the symmetry problem caused by the diagonal matrix, which is defined as follow:
\begin{itemize}
    \item \textbf{Symmetry Problem:}
    For triples in DistMult model satisfy the following equation:
    \begin{equation}
        \mathbf{h^{T}}diag(r)\mathbf{t}=\mathbf{t^{T}}diag(r)\mathbf{h}
    \end{equation}
\end{itemize}
In this situation, we call the relation is 'symmetry'. Nevertheless, the DistMult model acquiesces that every relation in KGs is self-symmetrical, which is quite different from the true situation.

To solve this problem, we share the tail entity embedding for each single DistMult model as Figure \ref{MDistMult} shows. For every single DistMult, the scoring function is as follow:
\begin{equation}
    f_{i}(h,r,t)=\mathbf{h}^{T}diag(r_{i})\mathbf{t_{s}},i=0,1,2,...,n
\end{equation}
where  $ i = {1,2,3,...,N}$, denotes the ith model. $n$ denotes the number of DistMult models, and $t_{s}$ is the shared tail entity embedding for all DistMult models. $r_{i}$ means the parameters in the diagonal matrix of the $i-th$ DistMult model.

By sharing the tail entity embedding of every single DistMult model, the symmetry problem can be solved and the shared tail embedding can be the shared part in each DistMult module. the proof is as follows:
\begin{itemize}
    \item \emph{Proof.} for a triple $(h,r,t)$ that input to the MDistMult, we have the scoring function:
    \begin{equation}
        S = \mathbf{h_{1}}diag(\mathbf{r_{1}})\mathbf{t_{1}} +...+\mathbf{h_{n}}diag(\mathbf{r_{n}})\mathbf{t_{1}}
    \end{equation}
    If the head and tail swap positions, we have the scoring function:
    \begin{equation}
        S^{'} = \mathbf{t_{1}}diag(\mathbf{r_{1}})\mathbf{h_{1}} +...+\mathbf{t_{n}}diag(\mathbf{r_{n}})\mathbf{h_{1}}
    \end{equation}
    Obviously:
    \begin{equation}
        S\neq S^{'}
    \end{equation}
\end{itemize}
We add all single DistMult model scoring function as $f_{all}$:
\begin{equation}
    f_{all}(h,r,t)=\sum_{i}^Nf_{i}(h,r,t)
\end{equation}
Different from the original DistMult model loss function, we use the softmax function to calculate the single DistMult model loss:
\begin{equation}
   \mathbf{P_{i}}(t|h,r) = \frac{exp(f_{i}(h,r,t))}{{\sum}_{t}exp(f_{i}(h,r,t))}
   \label{equ3}
\end{equation}

\begin{equation}
    loss_{i} = -log\mathbf{P_{i}}(t|h,r)
    \label{equ4}
\end{equation}
where $\mathbf{P_{i}}(t|h,r)$ is the probability of $t^{'}$ in $i-th$ DistMult model. In the triple (h,r,$t^{'}$), the $t^{'}$ is the missed entity that
need to be predicted. Furthermore, we construct the inverted triples $(t, r\_{reverse},  h)$ for each triple in KG. We will introduce the details in Section 4.2.
The whole MDistMult model scoring function for the same triple (h,r,$t^{'}$) has a similar loss function $loss_{all}$:
\begin{equation}
    loss_{all} = -log\mathbf{P_{all}}(t|h,r)
    \label{equ4}
\end{equation}
Considering the function of each single DistMult model scoring function and the overall scoring function in the MDistMult model, we combine both $loss_{i}$ and $loss_{all}$. The loss function of the MDistMult model is designed as follow eventually:
\begin{equation}
    Loss = loss_{all}+\sum_iloss_i
\end{equation}
In this way, our MDistMult model can be trained to obtain the overall optimal performance of multiple DistMult joint models.
\section{Experiments and Evaluation}\label{section4}
In this section, we evaluate our proposed MDistMult model on CCKS 2020 dataset and compared its performance with some baseline models. The results show that our MDistMult model has the best performance over all the other single models.
\subsection{Dataset}
We ran our experiment on CCKS 2020 COVID-19 Antiviral Drug Knowledge Graph (CADKG). The dataset of antiviral drugs includes four entities: drug, virus, viral protein and drug-protein, and four relations: effect, product, binding and interaction. The whole dataset includes 7844 entities, and we divide it into three parts: 36000 triples in the training set, 4000 triples in the valid set and 4256 triples in the test set. The details are shown in Table \ref{ADDS}. 
\begin{table}[h]\centering
\caption{Antiviral drug knowledge graph dataset}\label{ADDS}
\begin{tabular}{cccc}
\hline
                  & \textbf{Train set} & \textbf{Valid set} & \textbf{Test set} \\ \hline
\textbf{Entities} & \multicolumn{3}{c}{7844}                                    \\ \hline
\textbf{Relations} & \multicolumn{3}{c}{4}
\\\hline
\textbf{Triples}  & 36000              & 4000               & 4562              \\ \hline
\end{tabular}
\end{table}
\subsection{Baselines}
We compare our MDistMult model with multiple state-of-the-art KG embedding methods which mainly can be divided into four types as follows: Translation Models: TransE, TransR, TransH, TransD\cite{2015Knowledge}. Bilinear Methods: DistMult, ComplEx, SimplE\cite{2018SimplE}, TuckER\cite{2019TuckER}. Neural Network Model: ConvE.Rotation models: RotatE, QuatE. We report the results of these models on all evaluation metrics. The details are displayed in table \ref{tab1}.In addition, we explore the embedding effect on DistMult, RotatE and our MDistMult model, which is illustrated in figure \ref{dimension}.

\subsection{Data Augmentation}
Different from the usual experiment datasets for link prediction, such as FB15K\cite{bordes2013translating}, WN18RR\cite{2015Observed}, the amount of data in the CADKG is small comparatively. In order to get better model's performance and make the model better reflect the scoring function it designed. We designed the data augmentation method: for each triple $(h,r,t)$ in CADKG, we introduce inverted triple as $(t, r\_reverse, h)$. In this case, when we need to predict the triple $(?,r,t)$, we can equally predict the triple $(t, r\_reverse,?)$. Meanwhile, 
the amount of data in CADKG has doubled. The relations in CADKG are all unidirectional and are asymmetric. So the data augmentation method has no impact on all models results, and it makes the link prediction easily for all models.

\subsection{Evaluation Metrics}We use all the common link prediction metrics. The evaluation metrics include:  Mean Rank (MR) (an average rank of an answered entity overall test triplets ), Mean Reciprocal Rank (MRR, an average of the reciprocal rank of an answered entity overall test triplets), Hits at N (H@N). MRR is the main metric that we analyse since MRR is a more robust measure than mean rank, which would not change sharply when a single bad ranking appears. Yankai Lin et al.\cite{article} identified an issue that if the test set triples 
which were predicted existed in train set or valid set, it would appear at the top of the list, and we call it as \emph{raw} results. The \emph{raw} results will hide the real performance of the model in MRR since the top ranking has a big influence on MRR. So we remove the prediction results that appear in the train set or valid set, and we call this is \emph{filter} \cite{2017Knowledge}\cite{article}results. Our experiment results are all \emph{filter} results.
\subsection{Implementation}
We implemented all the compared models as well as our MDistMult model with the OpenKE\cite{han2018openke} project. And the parameters of these models are all well trained. So we can just replace the data to get all these results with the OpenKE project. For our proposed MDistMult model, the embedding dimension we set was 2000, and we used the Adam as our optimizer function and the learning rate was 0.0005. The Dropout was used too, the value of it was 0.5. We also used the L2 regularization, and the $\lambda$ of it was $1e^{-5}$. The batch size we set was 256. Considering the quantity and calculation of the MDistMult model comprehensively, we evaluated the case of the number of DistMult models $N$ equal to 2,3,4 respectively.
\begin{table}[h]
\caption{Experimental results}\label{tab1}
\centering
\begin{tabular}{l|ccccc}
\hline
\textbf{}                                                           
& \textbf{MRR} & \textbf{MR} & \textbf{H@1} & \textbf{H@3} & \textbf{H@10}        \\ \hline
TransE & 0.142 &487.59  & 0.040 & 0.188 & 0.334   \\
TransR & 0.104 &892  & 0.033 & 0.120 & 0.245  \\
TransH & 0.105 &773.39 & 0.027 & 0.126 & 0.263  \\
TransD& 0.103 &813.33 & 0.026 & 0.124 & 0.257 \\
DistMult & 0.090 & 1274.04& 0.038 & 0.098 & 0.197 \\
ComplEx & 0.073 & 1658.56& 0.027 & 0.066 & 0.172 \\
SimplE  & 0.084 & 747.28 & 0.015 & 0.093 & 0.238 \\
ConvE   & 0.180 & 758.17 & 0.102 &0.201	&0.340     \\
QuatE   & 0.105 & 620.31 & 0.023 & 0.118 & 0.280 \\
TuckER  & 0.096 & 627.00 & 0.045 & 0.098 & 0.193 \\
RotatE  & 0.200 & 521.50 & 0.113 & 0.233 & 0.369 \\ 
\hline
\textbf{MDistMult(N=2)} &0.243 &459.17 &0.150 &0.275 &0.429 \\
\textbf{MDistMult(N=3)} &0.244 &\textbf{455.35} &\textbf{0.152} &0.277 &\textbf{0.432}\\
\textbf{MDistMult(N=4)} &0.244 &458.88 &0.150 &\textbf{0.278} &0.430\\\hline
\end{tabular}
\end{table}
\subsection{Experiment Results}
Table \ref{tab1} shows the results of our experiments. It can be viewed that our proposed MDistMult model reach the best performance on all evaluation metrics no matter what $N$ is. The MRR achieves 24.4\%, the MR achieves 455.35, and H@1, H@3, H@10 are 0.152, 0.278, 0.432 respectively. Besides, our model had a huge improvement in all indicators even compared with the best performance model RotatE in the baseline. More importantly, we can find that the RotatE MRR is better than TransE, and exceed a lot, but the MR of RotatE is smaller than TransE. This means that the RotatE have good prediction results on top10 ranking, while on the whole CADKG, the TransE made a better prediction. MDistMult model doesn't have this problem, it has good performance on both the top 10 ranking predictions and the whole CADKG ranking prediction. This shows that our model has a balanced performance on the whole link prediction task.
\begin{figure}[H]
  \centering
  \includegraphics[width=0.5\textwidth]{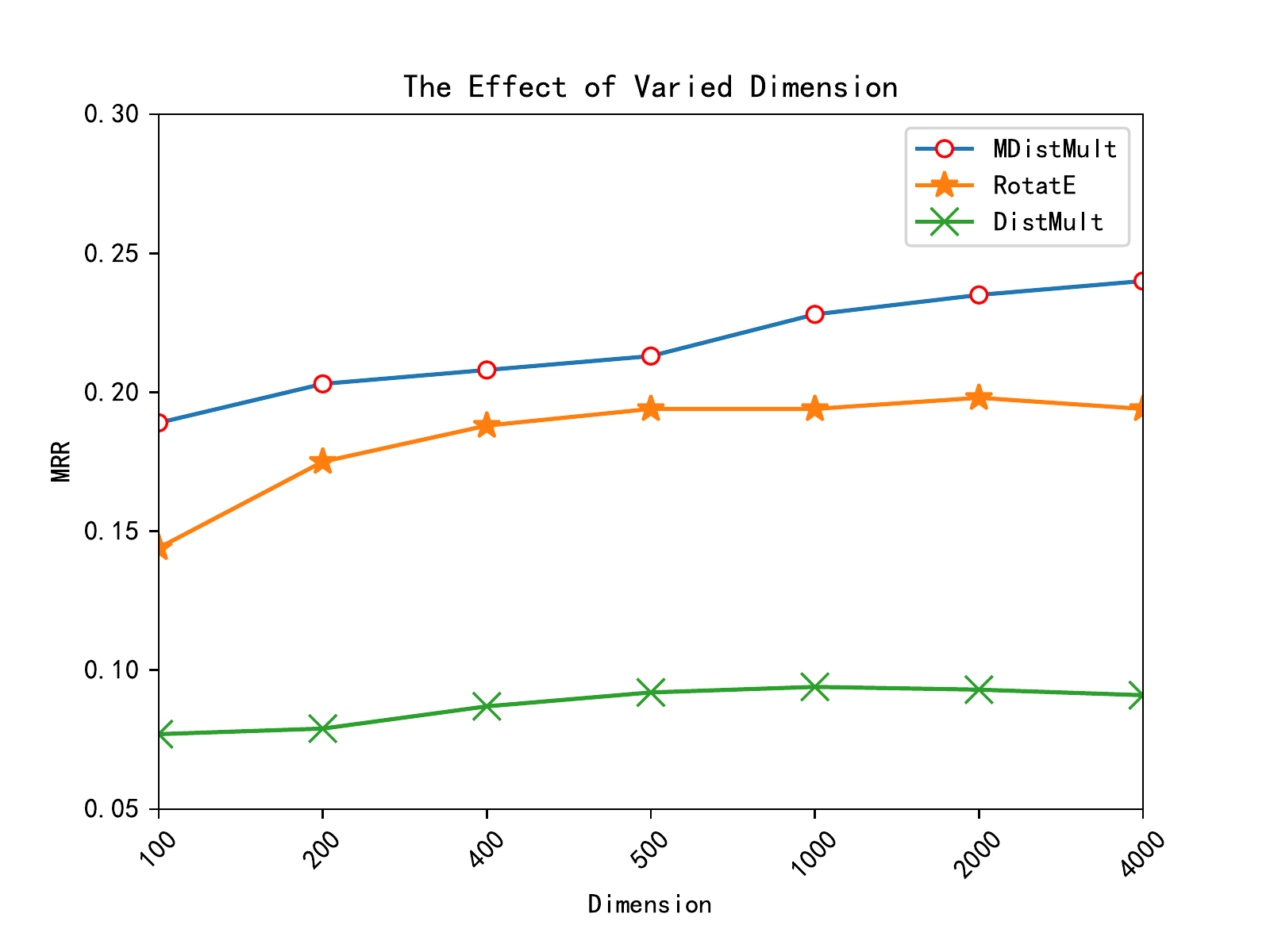}
  \caption{We show the change of MRR when the dimension of entity or relation increases. We can see a stable increasing trend of our proposed MDistMult model. Nevertheless, the MRR of RotatE and DistMult nearly stand at the same level when the dimension is over 500.}
  \label{dimension}
\end{figure}

\begin{figure}[H]
    \centering
    \hspace*{-1.5cm}
    \subfigure[Results on MRR]{
    \centering
    \begin{minipage}[h]{0.4\linewidth}
    \includegraphics[width=2in]{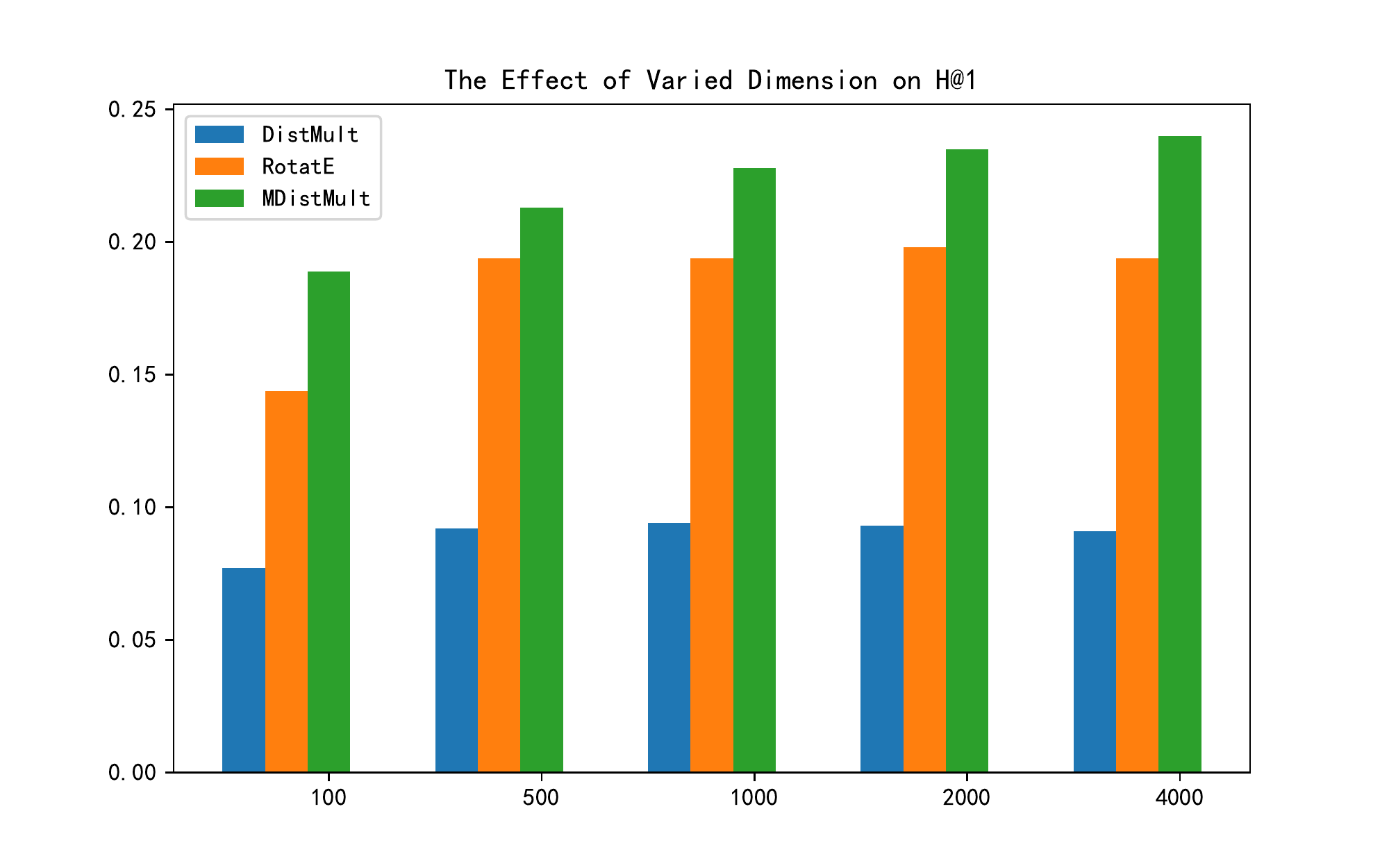}
    \label{MRR}
    \end{minipage}}
    \hspace*{0.5cm}
    \subfigure[Results on H@1]{
    \centering
    \begin{minipage}[h]{0.4\linewidth}
    \includegraphics[width=2in]{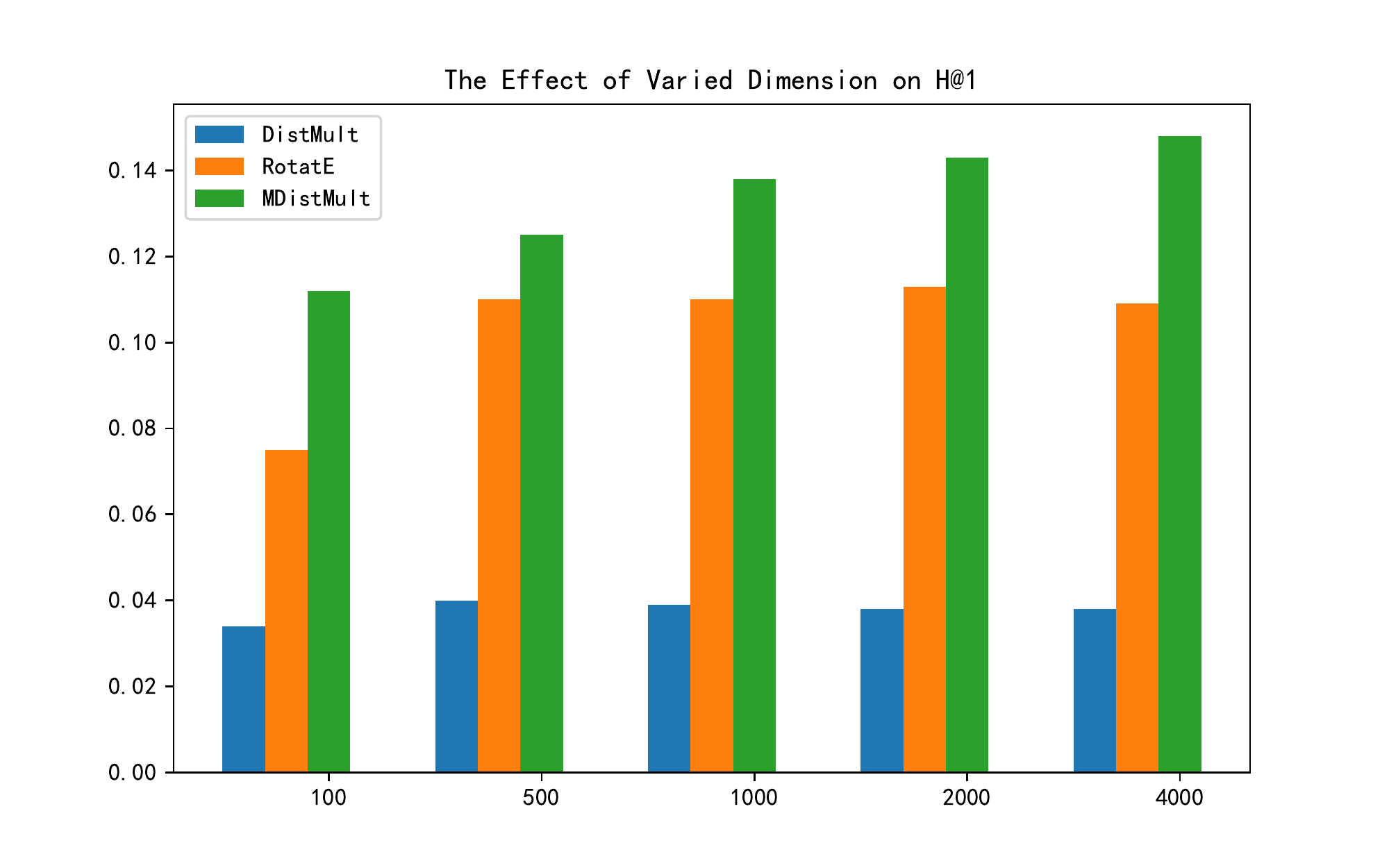}
    \label{H@1}
    \end{minipage}}
    
    \hspace*{-1.5cm}
    \subfigure[Results on H@3]{
    \centering
    \begin{minipage}[h]{0.4\linewidth}
    \includegraphics[width=2in]{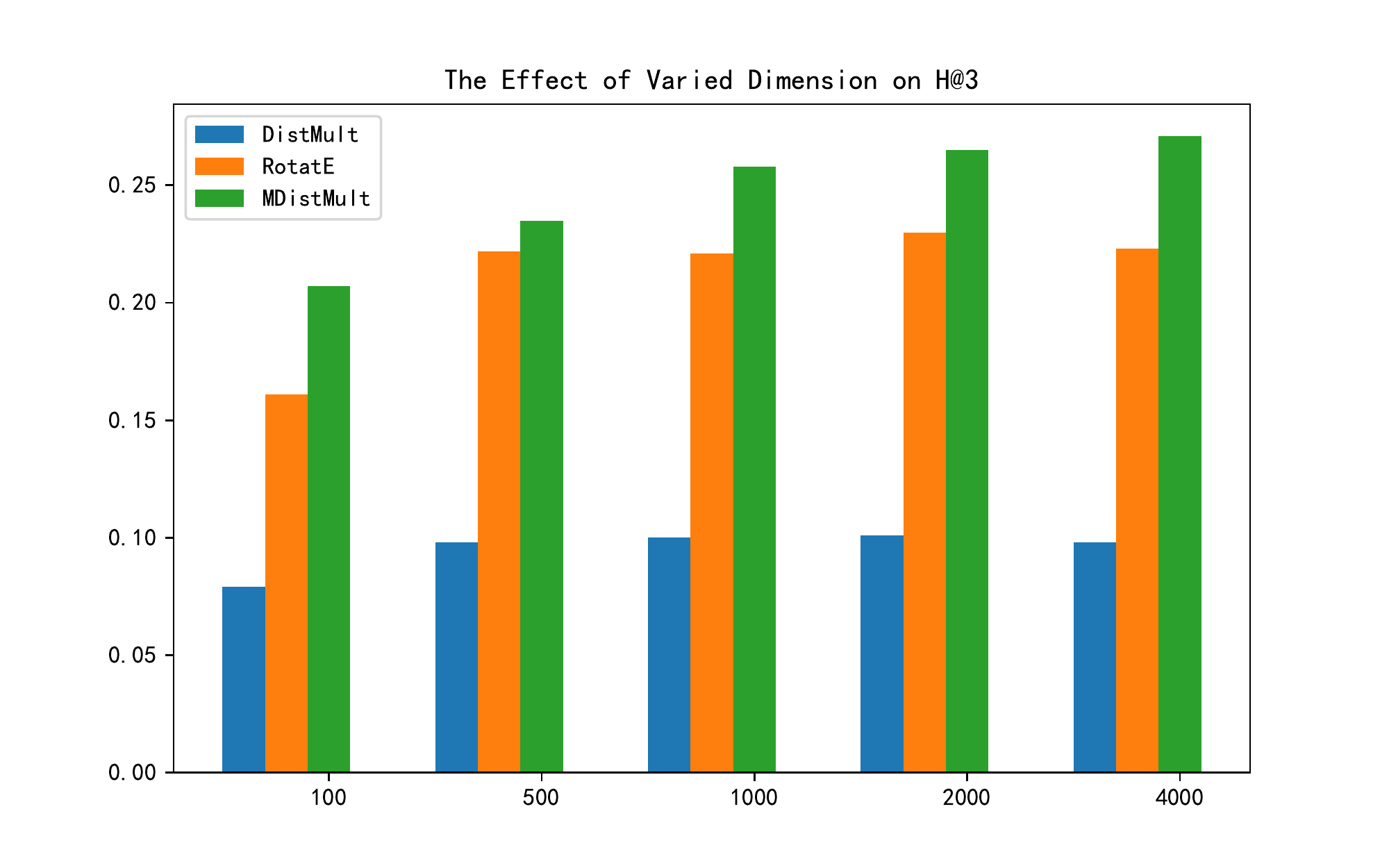}
    \label{H@3}
    \end{minipage}}
    \hspace*{0.5cm}
    \subfigure[Results on H@10]{
    \centering
    \begin{minipage}[h]{0.4\linewidth}
    \includegraphics[width=2in]{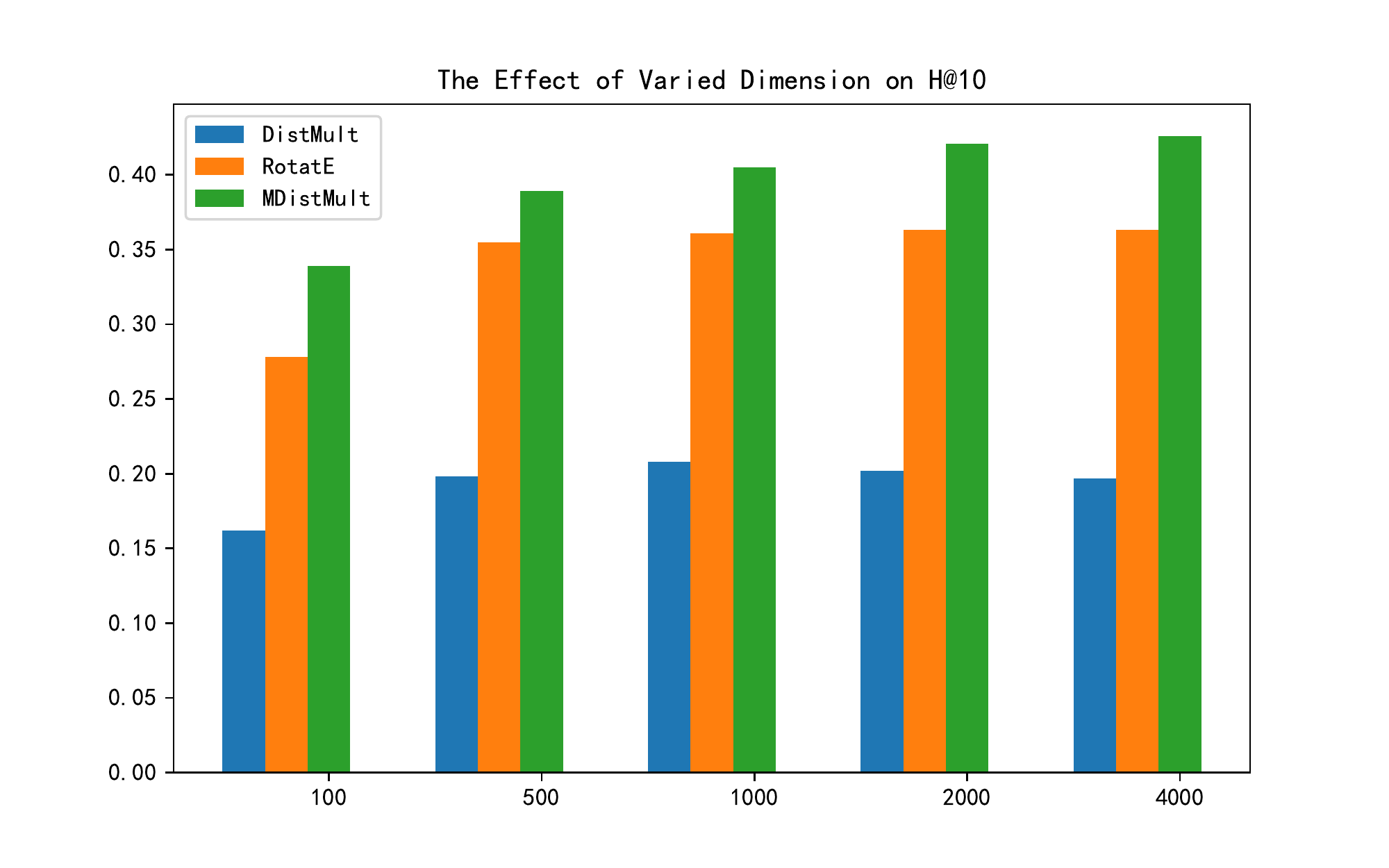}
    \label{H@10}
    \end{minipage}}
    \caption{More results on different metrics. The bar charts above illustrate that MDistMult has a steady increase on all metrics, while the other two models cannot increase when the dimension is big enough.}
    \label{alldim_result}
\end{figure}

For the MDistMult model, Table \ref{tab1} also shows that the performance is improved markedly when the amount of DistMult is up to 2, which also shows that the MDistMult model outperforms the RotatE model. When the number of DistMult modules in the MDitMult model increases, the performance can still have a significant improvement (when $N=3$, MR decreases around 4 points and H@1 increases 0.2\%), which can be explained both for MDistMult's advantages of overcoming the symmetry problem and extracting more features from the triples. The bold data above illustrate that our model achieves state-of-the-art performance compared with current KGE methods.

We further explore the impact of embedding changes on different models, which is shown in figure \ref{dimension}, and the $N$ is set to be $4$. The MRR of our proposed MDistMult is the largest no matter what dimension is at any point from 100 to 4000. At the same time, The MRR of ours soars sharply when the dimension changes from 500 to 1000, and then it increases smoothly. While the figures of RotatE and DistMult drop when the dimension is over 2000 and 1000 respectively. This means that our proposed model can extract more features due to the multiple scoring function design (By comparing MDistMult and RotatE and comparing MDistMult and DistMult respectively). And the dimension change experiment also shows the probability to achieve better performance by increasing the dimension when we employ a multiple scoring functions model.

At the same time, we explore the influence of dimension on more metrics of the DistMult, RotatE and MDistMult, which can be seen in figure \ref{alldim_result}. Our MDistMult model rises when the dimension is increased. And the MRR of our model is nearly three times than that of DistMult and this phenomenon can also be seen in \ref{H@1}, \ref{H@3}, \ref{H@10}. This is strong support for our design which aims to solve the \textbf{Symmetry Problem} which is described in section \ref{section3}. Apart from that, the RotatE model can obtain good performance just like our proposed model. However, it is clear that as the dimension increased, the gap between the RotatE and MDistMult is widening. We explain this phenomenon as the result that not only our model's climbing but also the RotatE's decrease when the dimension constantly increase. And this also proves that the design of multiple scoring functions can truly extract more features of triples to gain more information of knowledge graph and get better results. 

Another interesting thing we can learn from the little change of DistMult is that the symmetry problem has been a big limitation for DistMult. When we look into the table \ref{tab1}, we find the Translation based models have the same bad results, so the symmetry problem may be worse in some knowledge graphs of domain fields. The imbalanced distribution and the relation which is rarely Symmetrical in domain knowledge graphs may be the main reasons for the bad results. We know that the advantage of sampling makes RotatE reach powerful performance, which can be understood by the sharp increase of RotatE on all metrics when the dimension rises from $100$ to $500$. Compared with RotatE, our designed MDistMult doesn't rely on the sampling process. The increased performance of our model mainly depends on the design of multiple scoring functions. In addition, the dropped results of RotatE also mean that when the dimension is huge enough, the RotatE cannot extract more features from the triples and the sampling methods don't matter anymore.

\section{Conclusion and Future Work}
\label{section5}
In this paper, we proposed a new multiple scoring functions model named MDistMult, and it outperforms all the previous link prediction methods and has state-of-the-art results on the COVID-19 related dataset. In the future, we want to do more research on MDistMult and other multiple scoring functions models in several ways including 1-exploring the performance of MDistMult on more domain datasets, 2-attempting to build multiple scoring functions models for other single scoring function models on the link prediction task, 3- fusing different single scoring function models into multiple scoring functions models to get better performance.

\section{Acknowledgment}
We would like to thank the anonymous reviewers for their valuable and constructive comments. This work is supported by the National Key R\&D Program of China under Grant 2018YFC1604000.

\bibliographystyle{iopart-num.bst}
\bibliography{conference_101719.bbl}

\providecommand{\newblock}{}
\begin{thebibliography}{10}
\expandafter\ifx\csname url\endcsname\relax
  \def\url#1{{\tt #1}}\fi
\expandafter\ifx\csname urlprefix\endcsname\relax\def\urlprefix{URL }\fi
\providecommand{\eprint}[2][]{\url{#2}}

\bibitem{belinkov-glass-2019-analysis}
Belinkov Y and Glass J 2019 {\em Transactions of the Association for
  Computational Linguistics\/} {\bf 7} 49--72

\bibitem{michel2020covid}
Michel F, Gandon F, Ah-Kane V, Bobasheva A, Cabrio E, Corby O, Gazzotti R,
  Giboin A, Marro S, Mayer T {\em et~al.\/} 2020 {\em International Semantic
  Web Conference\/} (Springer) pp 294--310

\bibitem{domingo2020covid}
Domingo-Fern{\'a}ndez D, Baksi S, Schultz B, Gadiya Y, Karki R, Raschka T,
  Ebeling C, Hofmann-Apitius M and Kodamullil A~T 2021 {\em Bioinformatics\/}
  {\bf 37} 1332--1334

\bibitem{wise2020covid}
Wise C, Ioannidis V~N, Calvo M~R, Song X, Price G, Kulkarni N, Brand R, Bhatia
  P and Karypis G 2020 {\em arXiv preprint arXiv:2007.12731\/}

\bibitem{kaggle}
AI A 2020 {\em Allen Institute for Artificial Intelligence, https://www.
  kaggle. com/alleninstitute-for-ai/CORD-19-research-challenge\/}

\bibitem{bordes2013translating}
Bordes A, Usunier N, Garcia-Duran A, Weston J and Yakhnenko O 2013 {\em Neural
  Information Processing Systems (NIPS)\/} pp 1--9

\bibitem{2015Observed}
Toutanova K and Chen D 2015 {\em Proceedings of the 3rd workshop on continuous
  vector space models and their compositionality\/} pp 57--66

\bibitem{2014Knowledge}
Wang Z, Zhang J, Feng J and Chen Z 2014 {\em Proceedings of the AAAI Conference
  on Artificial Intelligence\/} vol~28

\bibitem{article}
Lin Y, Liu Z, Sun M, Liu Y and Zhu X 2015 {\em Proceedings of the AAAI
  Conference on Artificial Intelligence\/} vol~29

\bibitem{wang2018multi}
Wang Y, Gemulla R and Li H 2018 {\em Proceedings of the AAAI Conference on
  Artificial Intelligence\/} vol~32

\bibitem{nayyeri2019toward}
Nayyeri M, Xu C, Yaghoobzadeh Y, Yazdi H~S and Lehmann J 2019 {\em arXiv
  preprint arXiv:1909.00519\/}

\bibitem{2019RotatE}
Sun Z, Deng Z~H, Nie J~Y and Tang J 2019 {\em arXiv preprint
  arXiv:1902.10197\/}

\bibitem{ebisu2018toruse}
Ebisu T and Ichise R 2018 {\em Proceedings of the AAAI Conference on Artificial
  Intelligence\/} vol~32

\bibitem{2011A}
Nickel M, Tresp V and Kriegel H~P 2011 {\em Icml\/}

\bibitem{2014Embedding}
Yang B, Yih W~t, He X, Gao J and Deng L 2014 {\em arXiv preprint
  arXiv:1412.6575\/}

\bibitem{2016Complex}
Trouillon T, Welbl J, Riedel S, Gaussier {\'E} and Bouchard G 2016 {\em
  International Conference on Machine Learning\/} (PMLR) pp 2071--2080

\bibitem{2017Convolutional}
Dettmers T, Minervini P, Stenetorp P and Riedel S 2018 {\em Proceedings of the
  AAAI Conference on Artificial Intelligence\/} vol~32

\bibitem{2018A}
Vu T, Nguyen T~D, Nguyen D~Q, Phung D {\em et~al.\/} 2019 {\em Proceedings of
  the 2019 Conference of the North American Chapter of the Association for
  Computational Linguistics: Human Language Technologies, Volume 1 (Long and
  Short Papers)\/} pp 2180--2189

\bibitem{2019Quaternion}
Zhang S, Tay Y, Yao L and Liu Q 2019 {\em arXiv preprint arXiv:1904.10281\/}

\bibitem{2015Knowledge}
Ji G, He S, Xu L, Liu K and Zhao J 2015 {\em Proceedings of the 53rd annual
  meeting of the association for computational linguistics and the 7th
  international joint conference on natural language processing (volume 1: Long
  papers)\/} pp 687--696

\bibitem{2018SimplE}
Kazemi S~M and Poole D 2018 {\em arXiv preprint arXiv:1802.04868\/}

\bibitem{2019TuckER}
Bala{\v{z}}evi{\'c} I, Allen C and Hospedales T~M 2019 {\em arXiv preprint
  arXiv:1901.09590\/}

\bibitem{2017Knowledge}
Wang Q, Mao Z, Wang B and Guo L 2017 {\em IEEE Transactions on Knowledge and
  Data Engineering\/} {\bf 29} 2724--2743

\bibitem{han2018openke}
Han X, Cao S, Lv X, Lin Y, Liu Z, Sun M and Li J 2018 {\em Proceedings of the
  2018 conference on empirical methods in natural language processing: system
  demonstrations\/} pp 139--144

\end{thebibliography}

\end{document}